\documentclass{article}
\usepackage{spconf,amsmath,graphicx}
\usepackage{multirow}
\usepackage{booktabs}
\usepackage[colorlinks,linkcolor=red]{hyperref}
\usepackage{cleveref}

\title{Can Audio Captions Be Evaluated with Image Caption Metrics?}
\name{Zelin Zhou*, Zhiling Zhang*, Xuenan Xu*\thanks{*Equal contribution}, Zeyu Xie, Mengyue Wu$^\dagger$, Kenny Q. Zhu$^\dagger$ \thanks{$^\dagger$ Corresponding Author}}
\address{Author Affiliation(s)}
\address{Department of Computer Science and Engineering\\
	 Shanghai Jiao Tong University, Shanghai, China}

\begin{document}
\ninept
\maketitle
\begin{abstract}
Automated audio captioning aims at generating textual descriptions for an audio clip. To evaluate the quality of generated audio captions, previous works directly adopt image captioning metrics like SPICE and CIDEr, without justifying their suitability in this new domain, which may mislead the development of advanced models. This problem is still unstudied due to the lack of human judgment datasets on caption quality. Therefore, we firstly construct two evaluation benchmarks, \textit{AudioCaps-Eval} and \textit{Clotho-Eval}. They are established with pairwise comparison instead of absolute rating to achieve better inter-annotator agreement. Current metrics are found in poor correlation with human annotations on these datasets. To overcome their limitations, we propose a metric named \textit{FENSE}, where we combine the strength of Sentence-BERT in capturing similarity, and a novel Error Detector to penalize erroneous sentences for robustness. On the newly established benchmarks, FENSE outperforms current metrics by 14-25\% accuracy. 
\footnote{Code, data, and web demo available at: \url{https://github.com/blmoistawinde/fense}}
\end{abstract}
\begin{keywords}
Audio captioning, image captioning, caption evaluation, pre-trained model
\end{keywords}
\section{Introduction}
\label{sec:intro}
\textit{Automated audio captioning} \cite{drossos2017automated} is the task of automatically generating human-like content description of an audio signal using free text. Recent progress has been focused on the development of caption datasets \cite{drossos2020clotho, wu2019audio, kim2019audiocaps}, in which novel algorithms \cite{xu2021sjtu, mei2021encoder, xu2020crnn} are cultivated and fostered rapidly. However, little attention has been addressed on the automatic evaluation metrics. Current evaluations for audio caption directly adopt the metrics from the image captioning literature, including those for general Natural Language Generation (NLG) (\textit{e.g.} BLEU \cite{papineni2002bleu}, ROUGE \cite{lin2004rouge}, METEOR \cite{banerjee2005meteor}) or specifically for image captioning  (\textit{e.g.} CIDEr \cite{vedantam2015cider}, SPICE \cite{anderson2016spice}), without considering their generalizability to audio domain.

\begin{figure}[tbp]
  \centering
  \includegraphics[width=\linewidth]{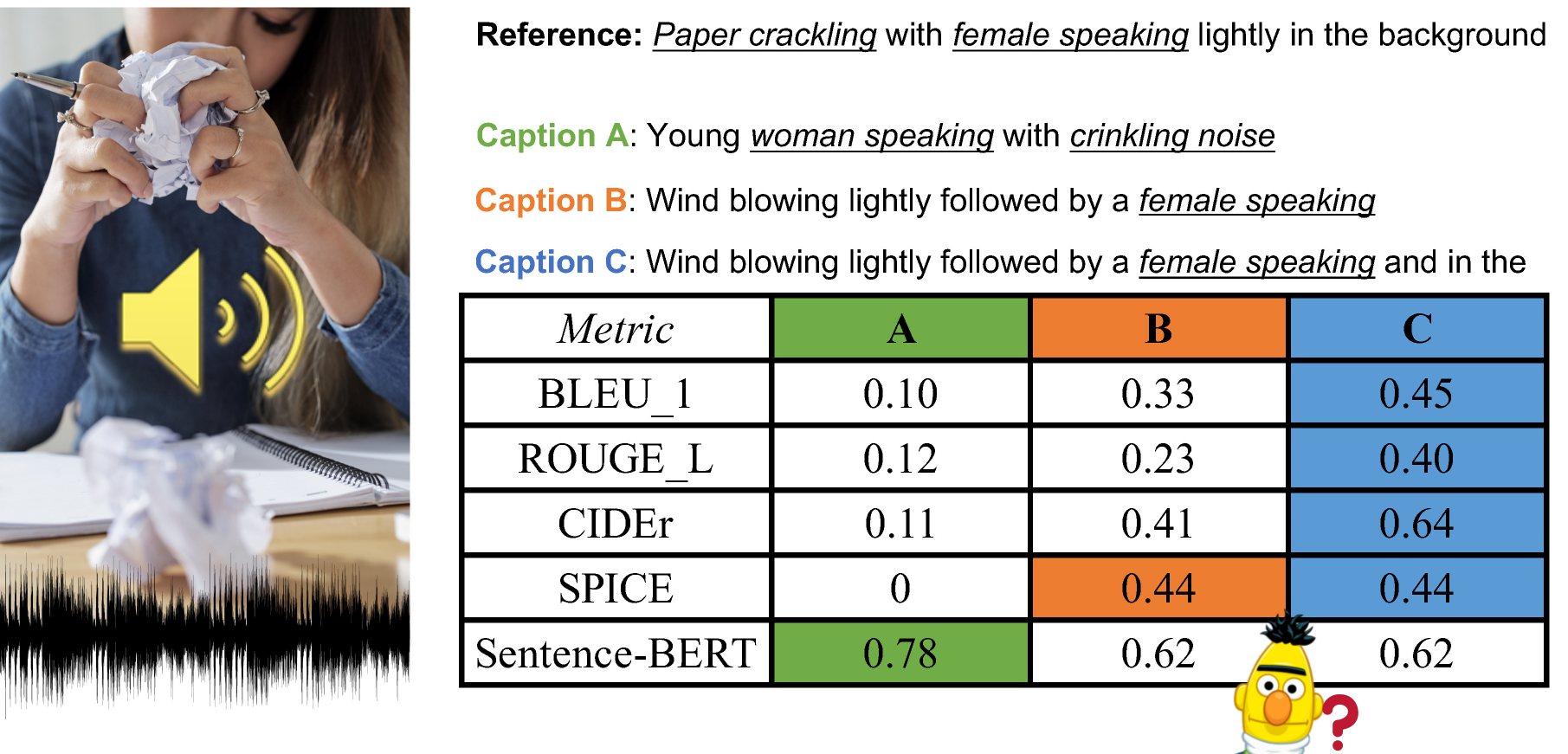}
  \caption{Column A, B and C shows the corresponding scores of Caption A, B and C given by different metrics. One cell is colored if the caption is favoured by a certain metric of this row. }
  \label{fig:example}
\end{figure}

In this work, we question: \textit{``Can audio captions be evaluated with image caption metrics?''}, since a biased evaluation may hinder the improvement of algorithms or even lead the journey to a skewed direction. In Fig \ref{fig:example}, we show that current metrics cannot properly evaluate audio captioning. On one hand, traditional metrics like BLEU may fail to capture semantic similarity beyond exact word matching \cite{novikova2017we, chaganty2018price}. Here, Caption A roughly paraphrases reference with all the sound events aligned, while Caption B partially aligned to reference with a misprediction. However, all the N-gram overlap based metrics give much higher scores to Caption B since it has more words in common with reference. 

On the other hand, the specificity in vision-focused captions may cause the failure of image caption specific metrics. 
In image captioning, attention has been drawn to the visible attributes of objects, along with their spatial relationships (\textit{e.g. A women in blue blouse sits at a table, with crinkled paper in her hand}). 
Conversely in audio captioning, importance has been attached to auditory properties of events as well as their temporal connections (\textit{e.g. Young women speaking with crinkling noise}). 
By this means, scene graph based metrics (\textit{e.g.} SPICE \cite{anderson2016spice}) are unlikely applicable to audio caption evaluation, in that object attributes and relations are seldom mentioned. 
It could be seen 
that SPICE even leaves a zero score on Caption A for its scene graph disjointness with reference, despite their semantic affinity. 

Moreover, \textit{fluency issues} like incomplete or inconsistent sentences are quite common in current generated captions. However, current metrics do not penalize, and sometimes even favour captions with these errors, so they can be easily gamed by systems exploiting the weakness. 
As shown in Fig \ref{fig:example}, Caption C attaches three meaningless words to Caption B, which further increases its N-gram overlap with reference. 
Scores indicate that all N-gram overlap based metrics are cheated by this trick, while the others are indifferent, which is also incompetence.  

Human evaluation is considered the golden standard in NLG tasks. However, by far there is no benchmark human judgments available for audio captioning.
It has certainly caused difficulty in conducting fair comparisons on evaluation metrics. 
Therefore, we annotated the first two human evaluation benchmarks for audio captioning, based on the widely accepted AudioCaps \cite{kim2019audiocaps} and Clotho \cite{drossos2020clotho} dataset. 
We experimented with two annotation protocols adopted from image captioning \cite{vedantam2015cider}: absolute rating and pairwise comparison. 
We find that absolute rating leads to poor inter-annotator agreement, and thus establish our final annotation based on pairwise comparison, resulting in the annotation of 3,421 pairs. 
We then benchmark current metrics, and find that metrics like BLEU$_1$ and SPICE perform no better than random guess on deciding a better generated caption in a pair.



To perform better audio caption evaluation, we explore the use of pre-trained model based metrics, including BERTScore \cite{zhang2019bertscore} and BLEURT \cite{sellam2020bleurt}, and witness significant advantage over current methods. 
We also repurpose Sentence-BERT \cite{reimers2019sentence}, which produces sentence embedding for similarity calculation and is mainly utilized for information retrieval, to perform caption evaluation (Fig \ref{fig:example}, Last Row), and it surprisingly achieved the best performance. 
However, even methods with enhanced capability for capturing semantic relatedness can still fail to penalize fluency issues properly.
We thus further propose \textit{Error Detector} to recognize erroneous sentences and penalize them accordingly. We refer to the combined metric of Sentence-BERT and fluency penalty as \textbf{F}luency \textbf{EN}hanced \textbf{S}entence-bert \textbf{E}valuation (FENSE), which significantly outperforms existing metrics.

In summary, our contributions are:
\begin{enumerate}
    \item We establish the first two benchmark datasets, \textit{AudioCaps-Eval} and \textit{Clotho-Eval}, for the comparison of audio caption evaluation metrics. 
    \item We propose FENSE, a metric consists of Sentence-BERT for similarity calculation and \textit{Error Detector} to penalize fluency issues for robustness.
    \item Results on the new benchmarks show that FENSE can outperform previous metrics in pairwise comparison accuracy by 14-25 points, and ablations suggest a significant contribution of both Sentence-BERT and Error Detector.
\end{enumerate}


\section{Benchmark Dataset Construction}
\label{sec:method}

\subsection{Audio Captioning Systems}
\label{ssec:caption_systems}
To generate candidate captions for evaluation, we involve multifaceted audio captioning systems: 
1) Nearest neighbor (NN) retrieval system. 
For a test audio, we retrieve its most similar training audio and take its annotation as a prediction. 
The similarity between two audio clips are measured by the cosine similarity of their embeddings, 
extracted by a pre-trained CNN~\cite{kong2020panns}.
2) Fully Connected (FC) input system.
An architecture similar to~\cite{wu2019audio}, where the input audio is first transformed to a vector and then decoded into a caption.
3) Attention (ATT) input system.
The same architecture as~\cite{xu2021sjtu}, where the input audio is first transformed into an embedding sequence and then decoded into a caption using temporal attention mechanism.
4) Reinforcement Learning (RL) system.
The same architecture as ATT system while the model is further fine-tuned using reinforcement learning~\cite{xu2021sjtu}, achieving the second place in the recent DCASE challenge evaluated by CIDEr and SPICE.

FC, ATT and RL systems are all sequence-to-sequence encoder-decoder frameworks.
They all use a 10-layer CNN as the encoder and a single-layer GRU as the decoder.
The detail structure can be found in~\cite{xu2021investigating}.
To increase the diversity of generated captions, for all systems except NN, we employ various decoding strategies, including greedy decoding, beam search, diverse beam search~\cite{vijayakumar2018diverse}.
In beam search, different temperatures are utilized.

\subsection{Data Annotation}
\label{ssec:data_annotation}
Based on captions generated by the aforementioned systems, we build the audio captioning evaluation dataset by collecting human judgments on captions.
We build our datasets based on two benchmark audio caption datasets: AudioCaps \cite{kim2019audiocaps} and Clotho \cite{drossos2020clotho}.
There are two protocols to collect human judgments in image captioning: absolute rating and pairwise comparison.
We first make a preliminary exploration on absolute rating protocol.
Given an audio clip and a caption, raters are asked to score the caption from 1 to 4 considering both its relevance to the audio and its fluency.
We randomly sample 100 audio-caption pairs.
Each audio-caption pair is rated by four different human raters.
However, Fleiss Kappa score~\cite{fleiss1971measuring} for different raters is only 0.18 on Clotho and 0.23 on AudioCaps, indicating poor inter-annotator agreement.
Annotators also report difficulty in rating the caption. 

Therefore, the second protocol, pairwise comparison, is adopted.
Given an audio clip and a pair of candidate captions, four raters are asked to choose which candidate describes the audio better in terms of description accuracy and fluency.
Raters are allowed to choose ``I'm not sure'' if they cannot distinguish which candidate is better.
Following \cite{vedantam2015cider}, we generate four pair groups, namely human-human correct (HC), human-human incorrect (HI), human-machine (HM) and machine-machine (MM).
HC contains two human annotations describing the same audio.
HI also contains two human annotations but one describes another randomly-picked audio.
HM is formed by a human annotation and a machine generated caption for the same audio.
MM is composed of two machine generated captions describing the same audio.
On both Clotho and AudioCaps, we randomly pick 250 audio clips and generate the four kinds of pairs.
For each audio clip, we form one HC pair, one HI pair, one HM pair and several MM pairs since the evaluation metrics are usually applied to compare captions generated by different machine systems.
The number of MM pairs for each audio clip range from one to four.

To avoid indistinguishable cases between similar pairs, we adopt a filtering strategy for pair generation.
We first generate all candidate pairs and calculate the similarity of each pair using Sentence-BERT embedding. 
Then we exclude candidate pairs with a similarity higher than 0.9.
We randomly sample pairs from the remaining captions to obtain the final generated pairs for comparison.
For HC and HM, we randomly sample the pairs.
For MM, we randomly sample four pairs if there are more than four candidate pairs, otherwise we use all the pairs left.
In case of no pairs left after filtering, we select the pair with the lowest similarity.
In this way, we finally obtain human judgments on 1,750 pairs on Clotho and 1,671 pairs on AudioCaps.
Higher inter-annotator agreement is achieved on both datasets, indicated by a Fleiss Kappa score of 0.48 and 0.33 respectively. We will refer to them as \textit{AudioCaps-Eval} and \textit{Clotho-Eval}. 
To our best knowledge, this is the first dataset of human judgments on the quality of audio captions.
We evaluate the effectiveness of different metrics using this dataset.

\section{Evaluation Metrics}
In this section we briefly introduce the automatic evaluation metrics investigated in this work.
Traditional NLG Metrics and Image Caption Metrics are the current adopted ones for audio caption. Pre-trained Model Based Metrics are the existent metrics utilized in other domains and proposed to use for audio caption. Error Detector is the newly customized metric for evaluating the vastly present fluency issues in generated captions.\\

\noindent\textbf{Traditional NLG Metrics} 
\quad   Traditional NLG metrics mainly rely on N-gram matching. \textbf{BLEU} \cite{papineni2002bleu} counts the exact N-gram matches between a candidate and its corresponding reference sentences and calculate the precision, while \textbf{ROUGE} \cite{lin2004rouge} is the recall-based counterpart of BLEU. A clear limitation for these methods is that even a subtle difference in the wording for expressing a similar meaning will be counted as an error. \textbf{METEOR} \cite{banerjee2005meteor} is proposed to alleviate the problem by supporting word stems, synonyms and simple paraphrases. However, similar words beyond such changes are still unhandled, and none of them are capable of capturing similarity based on contextualized semantics.




\noindent\textbf{Image Caption Metrics}
\quad   For the evaluation of Image captioning models, many methods have been proposed to further leverage the specific characteristics of image captions. Since image captions can be diverse even for the same picture, \textbf{CIDEr} \cite{vedantam2015cider} tries to capture the consensus among multiple annotators by rating a candidate by the mean TF-IDF similarity across reference captions. Moreover, since image captions usually describe the objects, attributes and relations depicted in the image, \textbf{SPICE} \cite{anderson2016spice} proposes to utilize the scene graph representation parsed from caption to capture such key concepts, and rate a candidate caption by the F1-score over scene graph tuples between a candidate and its references. 

These metrics have shown high correlation with human judgments for image captioning. However, some specific characteristics of image captions may not apply to audio captioning. For instance, the difficulty in identifying complex relations and attributes for an audio may pose challenges for the parse of scene graphs. They also suffer from the limitations of traditional metrics. 


 
\noindent\textbf{Pre-trained Model Based Metrics}
\label{sssec:bert_based_metrics}
\quad   Recently, pre-trained language models like BERT \cite{devlin2019bert} have been utilized for enhancing  the evaluation of NLG. Their ability to produce contextualized word representations can overcome the limitation of N-gram matching, and the large pre-training corpus may further facilitate their domain generalizability, which inspired us to investigate their performance for the evaluation of audio captioning. \textbf{BERTScore} \cite{zhang2019bertscore} leverages the contextualized word embeddings from the original BERT to calculate similarity on word-level, and aggregates them into sentence similarity with IDF weighting scheme. \textbf{BLEURT} \cite{sellam2020bleurt} follows the original BERT to pair the sentences as input and uses a linear layer on top of the [CLS] embedding to predict the score. They further finetune the BERT backbone specifically for evaluation.  

We also explore a novel use of \textbf{Sentence-BERT} \cite{reimers2019sentence}. It is a modification of BERT that uses siamese network structure to learn sentence embeddings so that their cosine similarity can reflect the semantic similarity. Although it was originally proposed for similarity search or clustering, we hypothesize that its capability of measuring semantic similarity may also benefit the evaluation of audio captioning. Therefore, we rate a candidate caption by its average cosine similarity with references according to Sentence-BERT embeddings.

\noindent\textbf{Error Detector}
\quad   Fluency issues like repeated events and incomplete sentences are prevalent in current audio captioning systems. However, few of the current evaluation metrics are able to take them into consideration, resulting in their overestimation of system quality. To mitigate this problem, we propose to use a separate error detector to penalize the scores given by other evaluation metrics when fluency issues are detected. 
We investigated the outputs of the captioning systems (\S \ref{ssec:caption_systems}), and found several typical types of fluency issues. We summarize them and give examples in Table \ref{tab:fluency_example}.

\begin{table}[h]
    \centering
    \small
    \begin{tabular}{l|l}
        \hline
        Type & Example \\
        \hline
        Incomplete Sentence & a woman is giving a speech and a \textit{(...)} \\
        Repeated Event  & music plays followed by \textit{music playing} \\
        Repeated Adverb & sheep bleats nearby several times \textit{nearby} \\
        Missing Conjunction & people speaking \textit{(and)} a train horn blows \\
        Missing Verb & food sizzles and a pan \textit{(verb)} \\
        \hline
    \end{tabular}
    \caption{Types and examples of fluency issues. We use parentheses to mark the missing information.}
    \label{tab:fluency_example}
\end{table}

To train a model for detecting such errors, we produce a synthetic training set by using rules to corrupt the correct captions into erroneous ones with such issues. Specifically, we use the captions in Clotho and AudioCaps training set as correct captions. We then tailor modifications for each error type to apply on the correct captions and label the modified ones as having the specific error type. For example, to produce incomplete captions, we append phrases that frequently appears at the tail of observed problematic examples like ``and'' , ``and a'', ``follow by'', etc. We empirically produce one error per caption 90\% of the time, and 2 errors otherwise. Then we mix the correct captions and corrupted captions into one set, and add an overall \textit{Error} label to samples with at least one error. Finally, we train a BERT model on this set for a multi-label classification of each error type and an overall \textit{Error} label.

To include the trained model for penalization, we take the model's predicted probability for \textit{Error}. If it exceeds a predefined threshold (0.9 in this work), we will divide the original score by 10. We observe improvement when combining the penalty with any metric, and obtain the best performance with Sentence-BERT (\S \ref{ssec:eval_two}). We thus propose FENSE as the combination of them. 


\section{Experiments}
\label{sec:experiments}

\begin{table*}[t]
\centering
    \begin{tabular}{l|c|c|c|c|c||c|c|c|c|c} \toprule
             Metrics     & \multicolumn{5}{c||}{AudioCaps-Eval} & \multicolumn{5}{c}{Clotho-Eval}   \\  \hline
          & HC    & HI   & HM   & MM   & Total & HC   & HI   & HM   & MM   & Total \\  \hline
    $\text{BLEU}_\text{1}$        & 58.6  & 90.3 & 77.4 & 50.3 & 62.4  & 51.0 & 90.6 & 65.5 & 50.3 & 59.0  \\
    $\text{BLEU}_\text{4}$        & 54.7  & 85.8 & 78.7 & 50.6 & 61.6  & 52.9 & 88.9 & 65.1 & 53.2 & 60.5  \\
    METEOR        & 66.0 & 96.4 & 90.0 & 60.1 & 71.7  & 54.8 & 93.0 & 74.6 & 57.8 & 65.4  \\
    $\text{ROUGE}_\text{L}$       & 61.1  & 91.5 & 82.8 & 52.1 & 64.9  & 56.2 & 90.6 & 69.4 & 50.7 & 60.5  \\  \hline
    CIDEr         & 56.2  & 96.0 & 90.4 & 61.2 & 71.0  & 51.4 & 91.8 & 70.3 & 56.0 & 63.2  \\  
    SPICE         & 50.2  & 83.8 & 77.8 & 49.1 & 59.7  & 44.3 & 84.4 & 65.5 & 48.9 & 56.3  \\  \hline
    BERTScore     & 60.6  & 97.6 & \textbf{92.9} & 65.0 & 74.3  & 57.1 & \textbf{95.5} & 70.3 & 61.3 & 67.5  \\
    BLEURT        & \textbf{77.3}  & 93.9 & 88.7 & 72.4 & 79.3  & 59.0 & 93.9 & 75.4 & 67.4 & 71.6  \\
    Sentence-BERT & 64.0  & \textbf{99.2} & 92.5 & 73.6 & 79.6  & 60.0 & \textbf{95.5} & 75.9 & 66.9 & 71.8  \\  \hline
    FENSE        & 64.5  & 98.4 & 91.6 & \textbf{84.6} & \textbf{85.3}  & \textbf{60.5} & 94.7 & \textbf{80.2} & \textbf{72.8} & \textbf{75.7} \\ \bottomrule
    \end{tabular}
    \caption{Benchmarking metrics for audio caption evaluation. Results are the correlation with human on pairwise comparisons.}
    \label{tab:all_results}
\end{table*}

\begin{table}[]
    \small
    \begin{tabular}{l|c|c|c|c|c|c} \toprule
                  & \multicolumn{3}{c|}{AudioCaps-Eval} & \multicolumn{3}{c}{Clotho-Eval} \\ \hline
                  & w/o  & w  & gain      & w/o & w     & gain      \\ \hline
    $\text{Bleu}_\text{1}$        & 62.4    & 75.3   &12.9      & 59           & 68.2       &9.2      \\
    SPICE         & 59.7    & 62.8 &3.1            & 56.3      & 62.1        &5.8        \\
    Sentence-BERT & 79.6    & \textbf{85.3}   &5.7          & 71.8            & \textbf{75.7}    &3.9         \\ \bottomrule
    \end{tabular}
    \caption{Ablation results of Error Detector in total accuracy. }
    \label{tab:fluency}
\end{table}

\subsection{Experimental Setup}
We conducted experiments on the two established datasets to compare different metrics' effectiveness on deciding a better caption out of two candidate sentences. For each pair of candidates, we compute all the metrics for two sentences and consider the one with a higher score as better according to the metric. A decision is considered correct if it agrees with human judgment. We calculate per-category accuracy by dividing correct predictions with total predictions within the categories (HC, HI, HM and MM), and merge them as a total score with micro average. Note that we exclude samples with equal number of opposite human judgments (excluding ``not sure'') so that the chosen samples can be considered distinguishable.

Since we may also include reference sentences in comparison, we only choose 4 out of 5 sentences from original references in the calculation of metrics. For an HC pair, we exclude each used reference itself from the five references. For an HI pair or HM pair, we eliminate the first human correct sentence from original references as their common new references. For an MM pair, we form the $C_5^4$ combinations selecting 4 out of 5 references as their common new references and calculate metrics on each of the 5 derivatives, then the five scores are averaged counting for the final score. 

To enable a fair comparison across the pre-trained model based metrics, we choose models of roughly equal size (usually small) and pre-trained on paraphrase tasks (if possible), which are BERTScore (\textit{t5-paraphraser}), BLEURT (\textit{BLEURT-tiny}) and Sentence-BERT (\textit{paraphrase-TinyBERT-L6-v2}). Note here BERTScore (\textit{t5-paraphraser}) is much larger than the other two models. BLEU$_1$, BLEU$_4$, ROUGE$_L$, METEOR, CIDEr and SPICE are implemented with tools provided by DCASE challenge.
\footnote{https://github.com/audio-captioning/caption-evaluation-tools}

\subsection{Standalone Evaluation of Error Detector}

To evaluate the performance of the proposed \textit{Error Detector}, we randomly sampled 723 machine generated captions, and annotated if the caption has a fluency issue. We then compare the model's prediction on the overall \textit{Error} class (threshold=0.9) with our annotations. The model achieved a precision of 96.8, recall of 76.4, and F1 of 85.4.

\subsection{Results on \textit{AudioCaps-Eval} and \textit{Clotho-Eval}}
\label{ssec:eval_two}

Table \ref{tab:all_results} shows our experimental results on AudioCaps-Eval and Clotho-Eval across different metrics. The results are consistent on two datasets for a common explanation. Intuitively, HC and MM pairs pose bigger challenges for all metrics, as they are either equally good or equally bad on describing the corresponding audio clip. Nonetheless, pre-trained model based metrics generally perform much better than other metrics, especially our proposed FENSE which achieves the best overall accuracy on both datasets.
BERTScore falls 4-5 percent behind other pre-trained model based metrics despite its biggest model size. On AudioCaps-Eval, N-gram exact match metrics (BLEU, ROUGE) perform 14-18 points worse than Sentence-BERT, while BLEU$_1$ and BLEU$_4$ struggle to exceed a random guess accuracy on MM pairs, showing their shortage differentiating two machine generated audio captions. 
METEOR and CIDEr attain a moderate performance gain beyond N-gram exact match metrics, however, they are still far behind Sentence-BERT. 
Not surprisingly, 
SPICE performs the worst, even unable to obtain 50\% accuracy on AudioCaps MM and Clotho-Eval HC \& MM, 
which strongly confirms our hypothesis that image captioning specified metrics cannot be adopted in audio caption evaluation.

Table \ref{tab:fluency} shows the overall accuracy of three representative metrics with and without sentence penalty of our proposed \textit{Error Detector} (The last row: FENSE). With \textit{Error Detector}, all the three metrics achieve extra performance gain, even for Sentence-BERT. 
This suggests that metrics leveraging either N-gram matching or contextual semantic similarity will fail in detecting \textit{fluency issues}, and \textit{Error Detector} can help mitigate such issues, thus setting a new SOTA. 

\begin{figure}[tbp]
  \centering
  \includegraphics[width=0.95\linewidth]{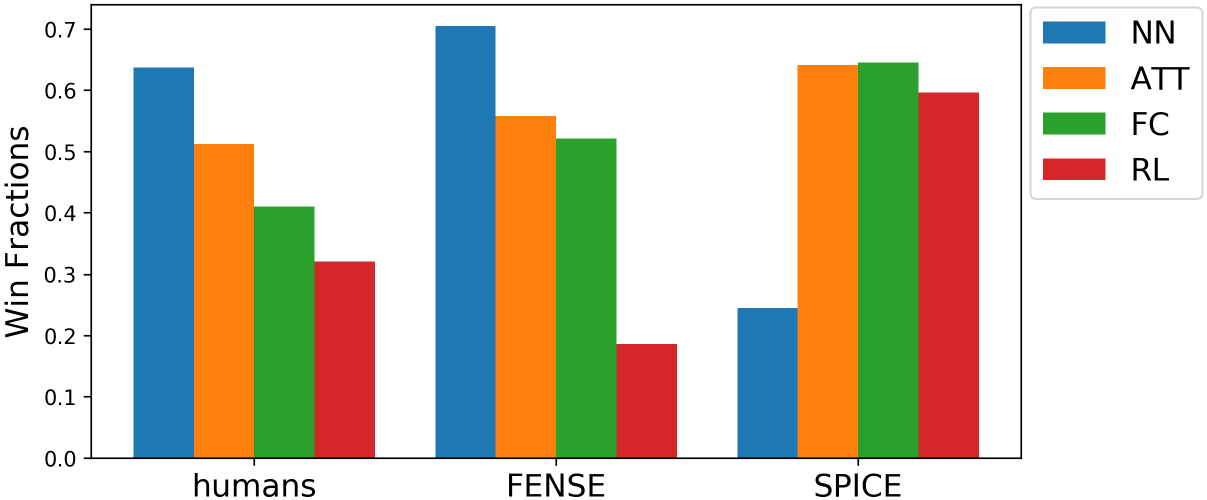}
  \caption{Illustration of judgments made by humans, FENSE and SPICE on Clotho-Eval. y-axis shows the win fractions of 4 audio captioning systems. }
  \label{fig:comparison}
\end{figure}

Integrating \textit{Error Detector} into Sentence-BERT metric, we get our proposed metric FENSE. Since table \ref{tab:all_results} and \ref{tab:fluency} have shown that FENSE significantly outperforms all the above metrics on the two evaluation datasets, we can use it to evaluate existing audio captioning system mentioned in \S \ref{ssec:caption_systems}. Fig \ref{fig:comparison} illustrates the judgments over four audio captioning systems made by humans, FENSE and SPICE on Clotho-Eval. Here we choose ATT, FC and RL whose beam search temperature equals to 0.5. On the y-axis we show the fraction of times one system wins another (rated better than another systems under this metric). We can see that NN is voted as the best system by both humans and FENSE, while it's the worst under SPICE. Humans and FENSE ranked ATT, FC and RL accordingly after NN. However, SPICE fails to reach their consensus again. This indicates the superiority of FENSE, and further confirms our claim that image captioning specified metrics like SPICE are inappropriate to be transferred into audio caption evaluation. 
\section{Conclusion} 
\label{sec:page}

In this work, we establish the first two human judgment datasets for audio caption evaluation, AudioCaps-Eval and Clotho-Eval, with pairwise comparison annotations. We benchmark commonly used metrics adopted from image captioning literature on the two datasets, and find their performance unsatisfying. We thus leverage Sentence-BERT for better estimation of semantic similarity and propose Error Detector to penalize sentences with fluency issues. The metric combining them two, named as FENSE, exhibits significant advantage.

Our findings suggest that the currently popular metrics for the research and competitions of audio captioning may not be a reliable measure of system quality, and better constitutes may be preferred. Although FENSE has achieved relatively good performance, we think that there is still sufficient room for improvement. Since semantic similarity does not equal to acoustic relevance \cite{Zhang2021EnrichingOW}, and only the former is covered in FENSE, better utilization of modality-specific knowledge may facilitate better evaluation as has been witnessed in image captioning \cite{jiang2019tiger}. 

\vfill
\pagebreak
\bibliographystyle{IEEEtran}
\bibliography{refs}

\end{document}